\documentclass[]{JHEP3}   

\JHEPspecialurl{http://jhep.sissa.it/JOURNAL/JHEP3.tar.gz}

\usepackage{graphicx}

\newcommand\fverb{\setbox\pippobox=\hbox\bgroup\verb}
\newcommand\fverbdo{\egroup\medskip\noindent%
            \fbox{\unhbox\pippobox}\ }
\newcommand\fverbit{\egroup\item[\fbox{\unhbox\pippobox}]}
\newbox\pippobox

\newcommand{\gsim}{\mathrel{\raisebox{-.6ex}{$\stackrel{\textstyle>}{\sim}$}}}
\newcommand{\be}{\begin{eqnarray}}
\newcommand{\ee}{\end{eqnarray}}

\def\sin{{\rm sin}}

\def\nota{\kern -1pt \not{\hbox{\kern 1pt $\scriptstyle{a}$}}}
\def\notb{\kern -1pt \not{\hbox{\kern 1pt $\scriptstyle{b}$}}}
\def\note{\kern -1pt \not{\hbox{\kern 1pt $\scriptstyle{e}$}}}
\def\notmu{\kern -1pt \not{\hbox{\kern 1pt $\scriptstyle{\mu}$}}}

\title{
\begin{flushright}
\normalsize{ FERMILAB-PUB-09-396-T\\FTUV 09-0811}
\end{flushright}
MINOS and CPT violating neutrinos}
\author{       {Gabriela Barenboim}\\
\normalsize\emph{Departament de F\'isica Te\`orica and IFIC, Universitat de
Val\`encia - CSIC}\\
\emph{Carrer Dr. Moliner 50, E-46100 Burjassot (Val\`encia), Spain}\\
Email: \email{gabriela.barenboim@uv.es}\\}

\author{\textbf{Joseph D. Lykken}\\
\normalsize\emph{Fermi National Accelerator Laboratory}\\
\emph{P.O. Box 500, Batavia, IL 60510, USA}\\
Email: \email{lykken@fnal.gov}}

\abstract{We review the status of $CPT$ violation in the neutrino
sector. Apart from LSND, current data favors three flavors of
light stable neutrinos and antineutrinos, with both halves of the
spectrum having one smaller mass splitting
and one larger mass splitting. Oscillation data for
the smaller splitting is consistent with $CPT$.
For the larger splitting, current data favor
an antineutrino mass-squared splitting that is an order of magnitude
larger than the corresponding neutrino splitting, with the corresponding
mixing angle less-than-maximal. This $CPT$-violating
spectrum is driven by recent results from MINOS, but is consistent
with other experiments if we ignore LSND.
We describe an analysis technique which, together with
MINOS running optimized for muon antineutrinos, should be
able to conclusively confirm the $CPT$-violating spectrum proposed
here, with as little as three times the current data set.
If confirmed,
the $CPT$-violating neutrino mass-squared difference would be an order of
magnitude less than the current most-stringent upper bound on
$CPT$ violation for quarks and charged leptons.
} 


\begin{document}


\section{Introduction}

All known particles are either self-conjugate under $CPT$ or have
$CPT$ conjugate ``antiparticles". In every case the antiparticle
partner is observed to have the same mass as the corresponding
particle, within experimental resolutions. These observations are
consistent with the description of all non-gravitational particle interactions
by local relativistic quantum field theory, where $CPT$ conservation
is a result of the intimate connection between Lorentz invariance,
locality, hermiticity, and the absence of operator ordering ambiguities.
For precisely this reason it is important to pursue increasingly rigorous
tests of $CPT$ invariance, and to extend our experimental constraints
to sectors previously beyond reach. 

In this regard neutrinos are especially interesting. Neutrinos have tiny
nonzero masses, suggesting that the neutrino mass generation mechanism
has novel features and that neutrinos communicate to a sector
of new physics whose effects on charged leptons and quarks are as yet
unobservable.
As demonstrated in the next section, the
current generation of neutrino
oscillation experiments are sensitive to $CPT$-violating effects
orders of magnitude smaller than what so far could have been
detected for charged leptons or quarks. There is both
theoretical and experimental motivation to pursue a rigorous
study of $CPT$ properties for neutrinos, keeping in mind that
$CPT$ violation may correlate with other exotic effects such as
Lorentz violation or quantum decoherence.

In this report we update \cite{Murayama:2000hm}-\cite{Kostelecky:2003xn}
the experimental constraints on $CPT$
violation for neutrinos, focusing on the case where other new
physics effects are subdominant to a $CPT$-violating difference
in neutrino/antineutrino mass spectra. As favored by the data we
also assume three flavors of light stable neutrinos and antineutrinos, 
both halves of the spectrum having  one smaller ``solar"
mass difference and one larger ``atmospheric" mass difference.
For the larger splitting we show that the global data set favors
an antineutrino mass-squared splitting that is an order of magnitude
larger than the corresponding neutrino splitting, as well as an
antineutrino mixing angle $\bar{\theta}_{23}$ that is less than maximal. This $CPT$-violating
spectrum is driven by recent results from MINOS \cite{newMINOS},
but is consistent with other experiments. 

We describe an analysis technique to confirm
or deny the best-fit $CPT$-violating hypothesis with future data. 
We advocate and demonstrate the use of the Neyman-Pearson hypothesis test \cite{James},
also known as the $\alpha$-$\beta$ test, generalized from ratios of
simple likelihoods to ratios of extended likelihoods with floating parameters.
This method has the advantage, for a given likelihood ratio, of
distinguishing between the test significance $\alpha$, the probability that $CPT$-conserving
masses and mixings are rejected even though they are in fact correct, 
and the power of the test $1-\beta$, where $\beta$ is the probability that
the $CPT$-violating solution is rejected even though it is
in fact correct. For a given future data set, one can require that the p-value
of the $CPT$-conserving hypothesis as
extracted from the likelihood ratio is less than some benchmark significance
$\alpha$ chosen according to one's theoretical prejudice about $CPT$ violation.

In advance of new data we can use Monte Carlo experiments to extract the value of $\beta$,
thus estimating the prospects for distinguishing $CPT$ violation in the
neutrino spectrum if it is in fact present. We examine these prospects for
the MINOS experiment. To be conservative, in maximizing the likelihoods
we do not float parameters defining the $CPT$-violating mass spectrum,
since this would tend to increase the maximum likelihood for the $CPT$-violating 
hypothesis even when it is wrong. We do however float 
experimental parameters related to the overall neutrino production rate
and the energy spectrum; floating these parameters
increases the maximum likelihoods for the incorrect hypotheses
while leaving the maximum likelihoods for the correct hypotheses
essentially unchanged, thus lessening the power of the
Neyman-Pearson test.

Even with this conservative approach, we demonstrate that
MINOS running optimized for muon antineutrinos should be
able to conclusively confirm the $CPT$-violating spectrum proposed
here, with as little as three times the current data set.

\section{$\mathbf{CPT}$ violation in the neutrino sector}

\subsection{$\mathbf{CPT}$ violation with and without Lorentz violation
or other exotic new physics}

The discovery of parity ($P$) violation in fundamental interactions
was a big surprise, especially considering that $P$ is an element of
the extended Lorentz group. As we now understand, it is possible to
violate $P$ in quantum field theory without compromising invariance
under the restricted Poincar\'e group that includes only 
proper orthochronous Lorentz transformations, i.e. Lorentz
transformations continuously connected to the identity.

For $CPT$, the connection to Lorentz invariance is even stronger.
As emphasized by Feynman, in a local description of quantum
field theory the Lorentz invariance of off-shell
amplitudes requires combining processes with propagation of both
off-shell states and $CPT$ conjugates of those states.
Going the other way, Greenberg has shown \cite{Greenberg:2002uu} that in quantum field theory
$CPT$-violating mass differences on-shell inevitably lead to Lorentz-breaking
effects off-shell, with consequences for both locality and operator-ordering
in quantum field theory.

Because of the initimate theoretical connection between $CPT$ and Lorentz
invariance, experimental searches for $CPT$ violation are related to
experimental tests of Lorentz invariance. In both cases the most
straightforward experimental approach is to look for departures
from the expected relativistic on-shell dispersion relations for particles and
antiparticles:
\be\label{eqn:dispersion-rels}
E^2 = {\vec{p}\,}^2 + m^2,\qquad   \bar{E}^2 = {\vec{\overline{p}}\,}^2 + \bar{m}^2  ,
\qquad \bar{m} = m \,,
\ee
where here and throughout a bar denotes a quantum number of a
$CPT$ conjugate state. This relationship suggests three experimentally
distinct scenarios:
\begin{itemize}
\item Detectable violations of Lorentz invariance in the dispersion relations
for some particles, but conserving $CPT$ to within experimental resolutions.
\item Detectable violations of Lorentz invariance in the dispersion relations
for some particles, accompanied also by detectable violations of $CPT$.
\item Detectable violations of $CPT$ in the dispersion relations
for some particles, but conserving Lorentz invariance to within experimental resolutions.
\end{itemize}
The first two scenarios are motivated by the possibility of a spontaneous
breaking of vacuum Lorentz invariance, perhaps related to new Planckian
physics such as space-time foam, superstrings, or extra dimensions \cite{Barenboim:2004wu}-\cite{Diaz:2009qk}.
The third scenario is
motivated by the possibility of non-local physics whose primary on-shell
effect may be $CPT$ violation \cite{Barenboim:2002tz}.

\FIGURE{
\centerline{\includegraphics[width=6.5 truein]{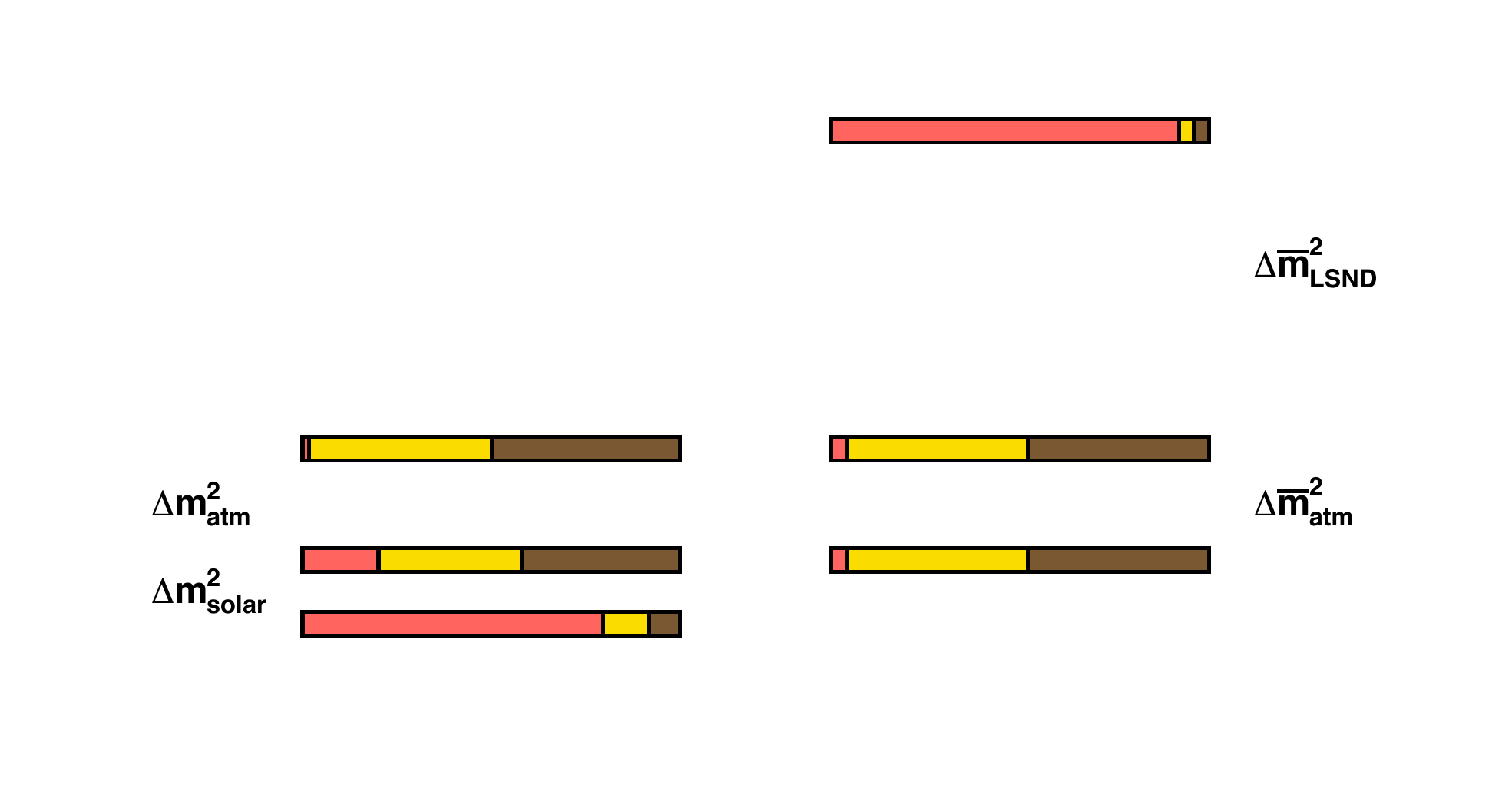}}
\caption
{\label{fig:one}
The $CPT$-violating neutrino spectrum
proposed in \cite{Barenboim:2001ac} as an explanation of LSND.
}
}

A further complication is that exotic new physics such 
as quantum decoherence \cite{Barenboim:2004wu}-\cite{Barenboim:2006xt}
or extra dimensions \cite{Hollenberg:2009tr}
may lead to baseline-dependent effects on neutrino oscillations with additional $CPT$-violating features
not captured by deviations from the expected dispersion relations.
Of course matter effects, though predicted by the Standard Model, are
also an example of baseline-dependent effects on neutrino oscillations with $CPT$-violating features.

For neutrino oscillation experiments there are thus effectively three kinds of tests
of $CPT$:
\begin{enumerate}
\item Searches for  Lorentz-violating effects in concert with $CPT$ violation. The current
best limits on this case are from the MINOS experiment \cite{MINOS:2008ij}; we will not
elaborate further on this scenario.
\item Searches for $CPT$-violating differences between neutrino and antineutrino mass spectra.
This is the main subject of our report.
\item Searches for inconsistencies in oscillation results that could signal
baseline-dependent new physics with possible ramifications for $CPT$.
Prospective limits are discussed in \cite{Mavromatos:2007hv,Sakharov:2009rn}
and \cite{Hollenberg:2009tr}.

\end{enumerate}

In the last case there is an important connection between $CP$ and $CPT$.
Even when $CPT$ is conserved, $CP$ violation in neutrino mixing allows the
possibility of differences between neutrino and antineutrino oscillation probabilities
in neutrino \textit{appearance} experiments:
\be
P(\nu_a \to \nu_b) \ne P(\bar{\nu}_a \to \bar{\nu}_b ) \; .
\ee
However, as shown in \cite{Barenboim:2002rv}, $CP$ violation without
$CPT$ violation cannot produce a neutrino-antineutrino discrepancy
in \textit{disappearance} experiments:
\be
P(\nu_a \to \nu_{\nota}) \ne P(\bar{\nu}_a \to \bar{\nu}_{\nota} ) \; .
\ee
A corollary of these results is that a neutrino-antineutrino oscillation
discrepancy arising from $CP$ violation without $CPT$ violation
requires at least two relevant mass splittings contributing to the 
oscillation, as occurs e.g. in
some $(3$$+$$2)$ sterile neutrino models \cite{Karagiorgi:2006jf}.

\FIGURE{
\centerline{\includegraphics[width=3.75 truein]{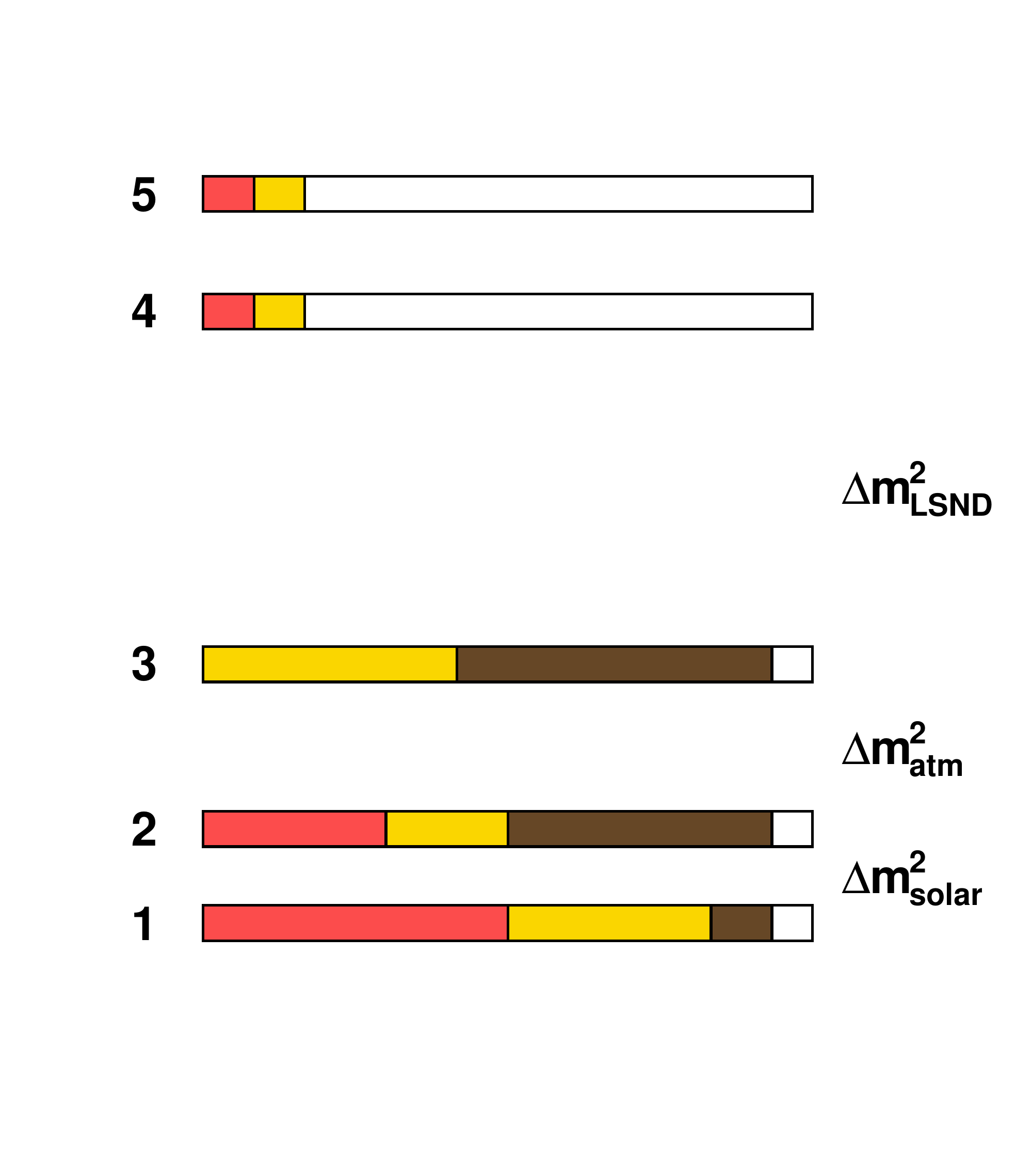}}
\caption
{\label{fig:two}
The $3$$+$$2$ $CPT$-conserving but potentially $CP$-violating
neutrino spectrum proposed in \cite{Karagiorgi:2006jf} attempting
to reconcile LSND, MiniBooNE, and other short-baseline oscillation
results.
}
}

\subsection{Comparing limits on $\mathbf{CPT}$ violation}

Assuming that the source of $CPT$ violation is a mass asymmetry in
the dispersion relations \ref{eqn:dispersion-rels}, the relevant
figure of merit in comparing different experimental limits on $CPT$-violation is
the mass-squared difference between a particle and its $CPT$ conjugate.

For quarks the most stringent experimental limit \cite{Amsler:2008zzb}
is from neutral kaons, whose mass-squared difference is constrained to
be less than 0.5 eV$^2$, or $\sim$ 0.1 eV$^2$ if we attribute the $CPT$
asymmetry to the constituent strange quarks.

For charged leptons, the most stringent constraint \cite{Amsler:2008zzb} is
from the upper limit on the electron-positron mass difference; this corresponds
to an upper bound on the mass-squared difference of approximately
$2\times 10^4$ eV$^2$.

The $CPT$-violating
best fit reported here corresponds to a difference of mass-squared
differences of only 0.02 eV$^2$. 
This means that for neutrinos the current generation of
oscillation experiments have sensitivities to potential $CPT$-violating
effects orders of magnitude smaller than the above limits. 
Note that Bahcall \textit{et al.} reached the same conclusion applying
different figures of merit \cite{Bahcall:2002ia} (see also \cite{Antusch:2008zj,Dighe:2008bu}).

Thus, contrary to what is sometimes implied in the literature, it is plausible that
$CPT$-violating mass differences would be detected first in the neutrino sector,
even if such effects have comparable magnitude in the quark and charged
lepton sectors. Furthermore, as noted already in the introduction, since
neutrinos appear to gain mass through a novel mechanism, it is also
plausible that $CPT$-violating mass differences are much larger for
neutrinos compared to the other sectors.

\section{$\mathbf{CPT}$-violating neutrino mass spectra}

The main constraints on the mass differences and mixings of neutrinos come
from the neutrino oscillation experiments \cite{chooz}-\cite{kamland}
summarized in Table \ref{expts_summary}.

\TABLE{
\caption{\label{expts_summary} Summary of current and past neutrino oscillation experiments.
The first column shows the principle oscillations that the experiment could in principle
observe; the second column indicates whether this constitutes an appearance
or disappearance experiment. The third column indicates the primary sensitivity,
either to ``solar" mass splittings, ``atmospheric" (atm), or ``short baseline" (SBL).}
\begin{tabular}{lccc}
\hline
CHOOZ \cite{chooz}, Bugey \cite{bugey}, Palo Verde\cite{paloverde} & $\bar{\nu}_e \to \bar{\nu}_{\note}$ & dis & SBL \\ \hline
CDHS \cite{cdhs}, CCFR \cite{ccfr}                   & $\nu_{\mu}\to\nu_{\notmu}$     & dis & SBL \\ \hline
NOMAD \cite{nomad}                                   &  $\nu_{\mu}\to\nu_e$       & app & SBL \\ \hline
LSND \cite{lsnd}, KARMEN \cite{karmen}                     & $\bar{\nu}_{\mu}\to \bar{\nu}_e$ & app & SBL \\ \hline
MiniBooNE \cite{miniboone}                            & $\nu_{\mu}\to \nu_e$ & app & SBL \\ 
                                                 & $\bar{\nu}_{\mu}\to \bar{\nu}_e$ &  &  \\ \hline
Super-Kamiokande \cite{superK}                                  &  $(\nu_{\mu}\to\nu_{\notmu}) + (\bar{\nu}_{\mu}\to \bar{\nu}_{\notmu})$ & dis & atm \\ \hline
K2K \cite{k2k}                                         &  $\nu_{\mu}\to\nu_{\notmu}$ & dis & atm \\ \hline
MINOS \cite{minos}                                    & $\nu_{\mu}\to\nu_{\notmu}$  & dis & atm \\
                                                 & $\bar{\nu}_{\mu}\to \bar{\nu}_{\notmu}$ & &  \\ \hline
SNO \cite{sno}                          & $\nu_e \to \nu_{\mu}$  & dis & solar \\ 
                                               & $\nu_e \to \nu_{\tau}$  &  &  \\  \hline
Borexino \cite{borexino}                      & $\nu_e \to \nu_{\note}$  & dis & solar \\ \hline
KamLAND \cite{kamland}                             & $\bar{\nu}_e \to \bar{\nu}_{\note}$ & dis & solar \\ 
\hline
\end{tabular}
}

Because we are interested in the possibility of $CPT$ violation, we will consider
the masses and mixings of the neutrino mass matrix as completely independent of the
the masses and mixings of the antineutrino mass matrix, and consider the experimental
constraints on each matrix separately. Because of the flavor sensitivity of the SNO results,
the active neutrino composition of the ``solar" neutrino oscillation is well-constrained.
The Super-Kamiokande data also have some flavor sensitivity; technically this 
measures the sum of the ``atmospheric" neutrino and antineutrino oscillations, but
in practice is mostly constraining for the neutrinos, which dominate over the
antineutrinos as cosmic ray secondaries in the relevant energy range. The Super-K
atmospheric neutrino data are bolstered by accelerator-based experiments K2K and MINOS,
which report muon neutrino disappearance consistent with the atmospheric mass
splitting and large mixing. The net
result \cite{GonzalezGarcia:2007ib} is that the active neutrino masses and mixings are required to closely resemble
the left half of the spectrum shown in Figure \ref{fig:one}, modulo the possibility
of inverting the solar-atmospheric hierarchy. The main question on the neutrino
side is whether there are small admixtures of one or more sterile neutrinos
in the three light mass eigenstates, but as yet there is no evidence for such mixings.

On the antineutrino side, the situation is less clear. KamLAND has reported 
an electron antineutrino disappearance signal consistent with an antineutrino mass splitting
and mixings equivalent to the ``solar" counterpart on the neutrino side.
LSND reported a $\bar{\nu}_\mu \to \bar{\nu}_e$ appearance signal 
consistent with an antineutrino mass-squared splitting $\sim 1$ eV$^2$.
MINOS has reported preliminary muon antineutrino disappearance results,
consistent with an antineutrino  mass-squared splitting that is roughly
the geometric mean of the KamLAND and LSND favored splittings.

Thus, even allowing for $CPT$ violation, oscillations between three active
antineutrino species cannot reconcile KamLAND, MINOS and LSND simultaneously.
The $CPT$-violating spectrum shown in Figure \ref{fig:one}, proposed 
in \cite{Barenboim:2001ac} to accommodate solar, atmospheric, and LSND
splittings with only three active flavors, was conclusively excluded by
KamLAND \cite{Barenboim:2002ah}. 

Without resorting to new baseline-dependent exotic physics, this leaves two
possibilities:
\begin{itemize}
\item {\bf Case (i)} The LSND results are incorrect.
\item {\bf Case (ii)}
The LSND results are correct, but the corresponding short-baseline
(SBL) oscillation involves mixing with one or more species of sterile neutrinos.
\end{itemize}

In the second case one may question whether it is even necessary to resort
to a $CPT$-violating mass spectrum, since the addition of sterile neutrinos
adds new parameters that potentially loosen up the experimental constraints.
Several recent analyses \cite{Karagiorgi:2006jf},\cite{Maltoni:2007zf}-\cite{Karagiorgi:2009nb}
have looked at this question in detail, using global fits that (in the case of \cite{Karagiorgi:2009nb})
include the latest MiniBooNE data. The conclusion is that $CPT$-conserving
sterile neutrino scenarios, even allowing for the possibility of large $CP$ violation
in the case of two or more sterile species, cannot avoid at least a 3 $\sigma$
discrepancy among different experimental data sets, with the largest tension
between the SBL appearance experiments and the SBL disappearance
experiments.

Thus Case (ii) requires either that we disregard some oscillation results
other then LSND, or that we again resort to a $CPT$-violating spectrum.
Thus for example one could develop $CPT$-violating versions of the
$3$$+$$2$ spectrum shown in Figure \ref{fig:two}. We will not pursue
this possibility further here, since it is already under 
investigation elsewhere \cite{mit}.

The remainder of this paper is devoted to Case (i): we disregard the LSND signal,
and explore to what extent a $CPT$-violating neutrino spectrum is allowed, or
even favored, by the remaining global data set. For simplicity we will assume
that the situation is not further complicated by sterile neutrinos or baseline-dependent
exotic physics, though of course both are possible.

With these assumptions there is qualitatively only one $CPT$-violating neutrino mass
spectrum candidate, shown in Figure \ref{fig:three} and first discussed in \cite{Barenboim:2002ah}.
To be more precise there are
four candidate spectra, since we can invert the hierarchy on either the neutrino
or antineutrino side independently, but existing data is insensitive to these choices,
with the exception of the neutrino observations from supernova SN1987A \cite{Murayama:2000hm}.

As discussed above, the neutrino side of the spectrum in Figure \ref{fig:three}
is completely constrained by data. On the antineutrino side, the smaller
``solar'' mass splitting is necessary to accommodate the KamLAND signal;
a $CPT$-violating variation of this splitting and the related mixings is still allowed
at about the 5\% level. The larger mass splitting has to accommodate
the antineutrino disappearance signals from MINOS and Super-K, 
the null appearance results from KARMEN and MiniBooNE, as well as the null
disappearance results from other SBL oscillation experiments.

\FIGURE{
\centerline{\includegraphics[width=6.5 truein]{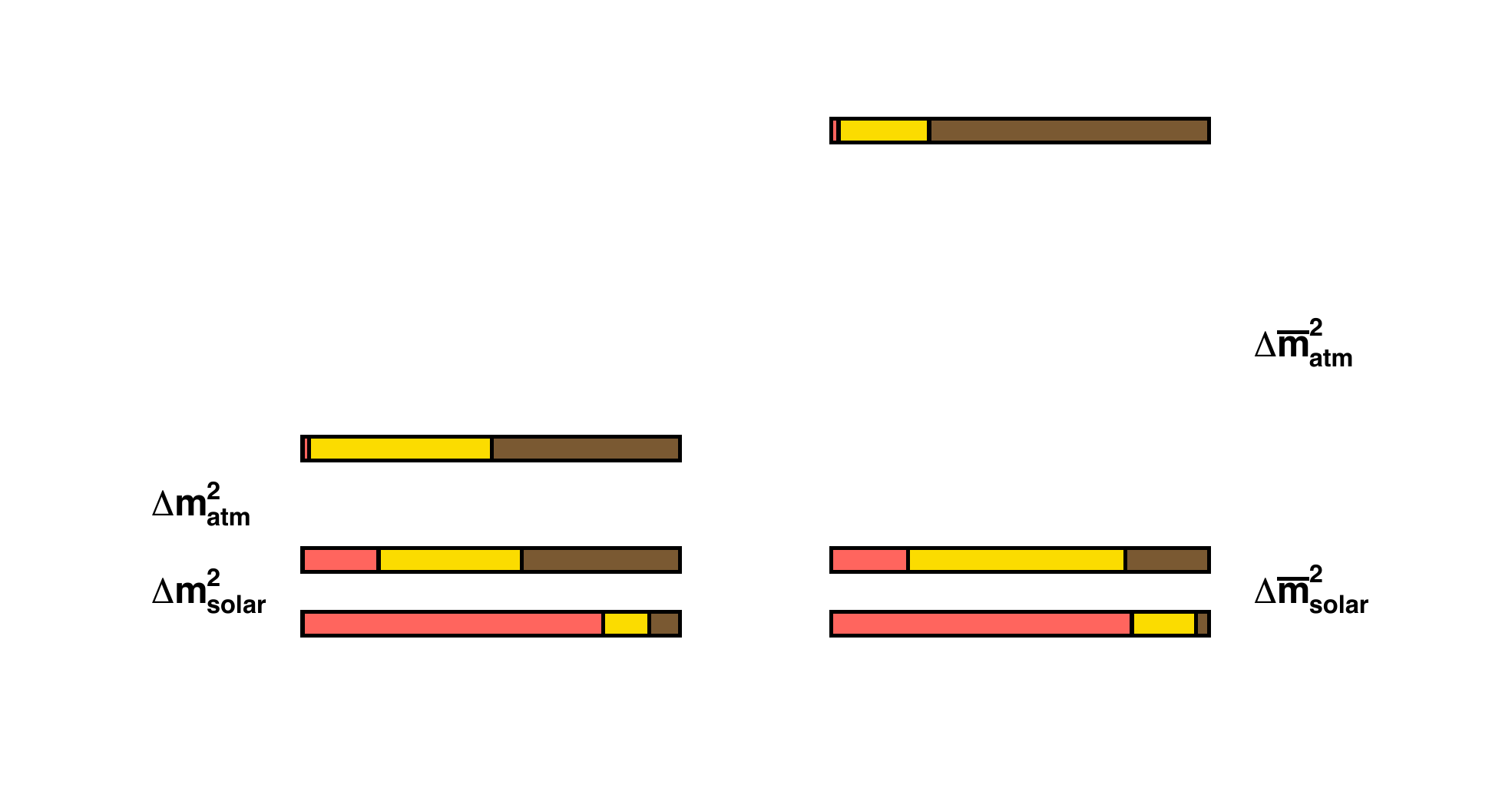}}
\caption
{\label{fig:three}
The best-fit $CPT$-violating neutrino spectrum obtained from our analysis.
}
}

\section{Constraining the antineutrino spectrum}

To make detailed contact with the experimental results we first introduce
the neutrino survival and transition probabilities given by
\be
P(\nu_\alpha \rightarrow \nu_\beta) = \delta_{\alpha \beta} -
 4 \sum_{i>j=1 }^3 U_{\alpha i} U_{\beta i} U_{\alpha j} 
U_{\beta j}
\;\sin^2 \left[ \frac{\Delta m_{ij}^2 L}{4 E} \right]
\label{pro}
\ee
for neutrinos and
\be
P(\overline{\nu}_\alpha \rightarrow \overline{\nu}_\beta) = 
\delta_{\alpha \beta} -
 4 \sum_{i>j=1 }^3 \overline{U}_{\alpha i} \overline{U}_{\beta i} 
\overline{U}_{\alpha j} \overline{U}_{\beta j}
\;\sin^2 \left[ \frac{\Delta \overline{m}_{ij}^2 L}{4 E} \right]
\label{apro}
\ee
for antineutrinos.
The matrix $U=\left\{ U_{\alpha i}\right\}$  
($\overline{U}=\left\{ \overline{U}_{\alpha i}\right\}$)
describes the weak
interaction neutrino (antineutrino) states, $\nu_\alpha$, in terms of the 
neutrino (antineutrino) mass eigenstates,
$\nu_i$. That is,
\be
\nu_\alpha = \sum_i U_{\alpha i} \nu_i \;\;\;\;\;\; \mbox{and} \;\;\;\;\;\;
\overline{\nu}_\alpha = \sum_i \overline{U}_{\alpha i} \overline{\nu}_i
\; ,
\ee
where we have ignored the possible $CP$ phases. The matrices can be
parametrized as follows:
\be
U=\pmatrix{c_{12} c_{13} & s_{12}c_{13} & s_{13} \cr
       -s_{12}c_{23} - c_{12}s_{23}s_{13} & c_{12}c_{23}- s_{12}s_{23}s_{13} & s_{23}c_{13} 
\cr
        s_{12}s_{23} - c_{12}c_{23}s_{13} & - c_{12}s_{23} - s_{12}c_{23}s_{13} 
&c_{23}c_{13}}
\ee
and similarly for $\overline{U}$.
In Eq.~(\ref{pro}) $L$ denotes the distance 
between the neutrino source and the detector, and
$E$ is the lab energy of the neutrino.

We use the notation
$\Delta m_{\rm solar}^2 = \Delta m_{12}^2$, $\Delta m_{\rm atm}^2 = \Delta m_{13}^2$ 
to denote the smaller and larger mass-squared splittings on the neutrino side,
and 
$\Delta \bar{m}_{\rm solar}^2 = \Delta \bar{m}_{12}^2$, $\Delta \bar{m}_{\rm atm}^2 = \Delta \bar{m}_{13}^2$ 
for the antineutrinos.

SBL reactor experiments give important constraints on the antineutrino spectrum.
Their results indicate \cite{chooz, bugey, paloverde}
that electron antineutrinos produced
in reactors remain electron antineutrinos on short baselines. 
Because of the short baselines we can ignore
the smallest (``solar'')  antineutrino mass difference and average the other two; 
the survival
probability can be expressed as
\be
P(\overline{\nu}_e \rightarrow \overline{\nu}_e) = 
1 - 2 \overline{U}_{e3}^2 (1- \overline{U}_{e3}^2) \; .
\ee
Thus, even for rather large antineutrino mass differences, the
survival probability will be close to one if $\overline{U}_{e3}$ is either
almost one or almost zero. Physically this means that we
can choose between having almost all the antielectron flavor
in the heavy state (which really means the furthest away state since we can
invert the spectrum) or alternatively
leave this state with almost no antielectron flavor. 
The first possibility was depicted in the Figure \ref{fig:one},
while the second is realized in Figure \ref{fig:three}.

MINOS and Super-Kamiokande constrain both the
larger antineutrino mass-squared difference
$\Delta\bar{m}_{\rm atm}^2$ and the antineutrino mixing
angle $\bar{\theta}_{23}$. KamLAND constrains mostly
the smaller antineutrino mass-squared difference
$\Delta \bar{m}_{\rm solar}^2$.

We have performed a $\chi^2$ fit of the 
antineutrino spectrum (assuming three active flavors only)
using the data from MINOS, Super-Kamiokande, KamLAND,
and CHOOZ.
The best-fit result is shown in Figure \ref{fig:three}.
The $CPT$-violating features are encapsulated in:
\be
\Delta\bar{m}_{\rm atm}^2 = 0.02{\rm\ eV}^2\;, \quad
{\rm sin}\,\bar{\theta}_{23} = 0.407 \; ,
\ee
compared to the global-fit neutrino spectrum values
\be
\Delta\bar{m}_{\rm atm}^2 = 0.0025{\rm\ eV}^2\;, \quad
{\rm sin}\,\theta_{23} = 0.707 \; .
\ee

This $CPT$ violation is driven by the MINOS results; indeed
our best-fit values for $\Delta\bar{m}_{\rm atm}^2$  and
sin$^22\bar{\theta}_{23}$ are close to those reported by
MINOS fitting their data alone.

The overall quality of our fit is good, with a $\chi^2$ per
degree of freedom of 0.98. As seen in 
Figure \ref{fig:four}, the $\chi^2$ deviation as a function of
the single variable $\Delta\bar{m}_{\rm atm}^2$ has a
clearly defined minimum. We note however that this is
only the case when $\bar{\theta}_{23}$ is allowed to
float in the fit; if $\bar{\theta}_{23}$ were fixed to maximal
mixing, then the chi-squared distribution in $\Delta\bar{m}_{\rm atm}^2$
would be rather flat.

\FIGURE{
\centerline{\includegraphics[width=6 truein]{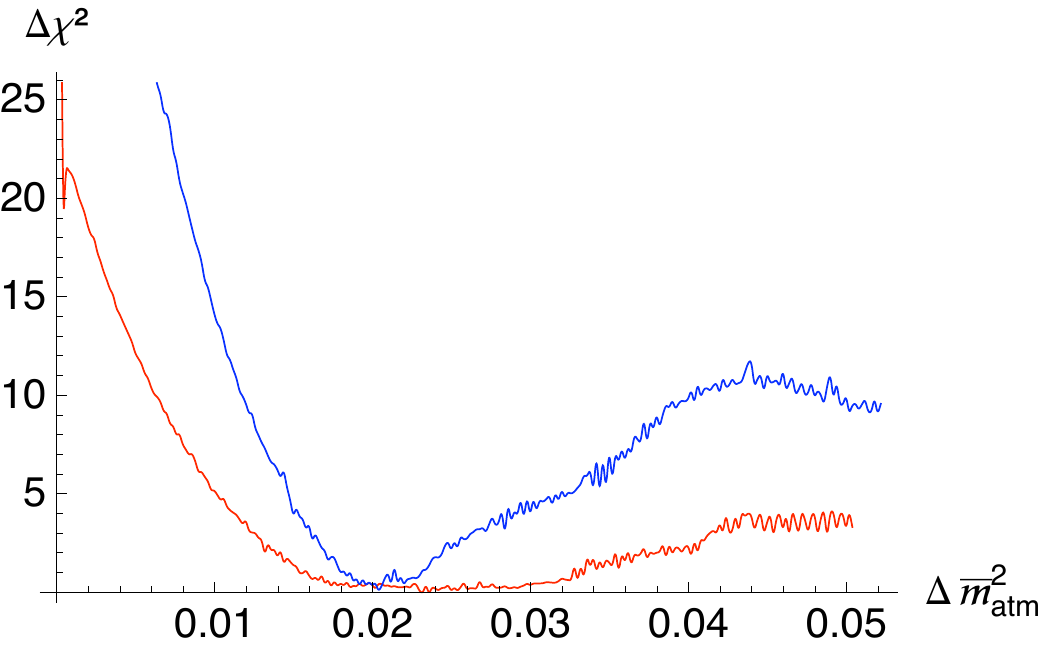}}
\caption
{\label{fig:four}
The relative $\chi^2$ deviation of the $CPT$-violating fit to data from MINOS,
Super-Kamiokande, KamLAND, and CHOOZ,
as a function of the single parameter $\Delta\bar{m}_{\rm atm}^2$
(in eV$^2$),
with ${\rm sin}\,\bar{\theta}_{23} = 0.407$. The lower (red)
curve excludes the MINOS muon antineutrino data,
while the upper (blue) curve includes it.
}
}

\section{Discussion and future prospects}

The MINOS muon antineutrino disappearance results should be regarded
as preliminary. They are from data runs with the target optimized for
neutrinos, introducing more complicated systematics for the antineutrinos,
and poorer statistics (42 events observed at the far detector).
This situation will improve dramatically with results from MINOS
running optimized for antineutrinos, scheduled to begin soon.

Our fit shows that large, order-of-magnitude $CPT$ violation
in the neutrino sector is still a viable possibility. Making the further assumptions
that the $CPT$ violation is (approximately) baseline-independent and does
not have a strong dependence on sterile mixing, a unique $CPT$-violating
mass and mixing pattern is selected, up to the four-fold ambiguity of
inverting the neutrino and/or antineutrino hierarchies.

The most timely question is whether better data in the near
future from MINOS could provide compelling evidence for neutrino
$CPT$ violation. To address this question, we have performed
toy muon antineutrino disappearance experiments, using the
survival probability obtained either from a $CPT$-conserving spectrum
or from our best-fit $CPT$-violating
spectrum. We use the reconstructed muon antineutrino energy spectrum
reported by MINOS, but to add some realism we allow a one-parameter
distortion of the energy spectrum, and float this parameter in the fit.
We also float $N_0$, the mean expected number of neutrinos detected 
in the MINOS far detector in the absence of oscillations; while this
number is estimated in the experiment, it is subject to significant
systematic uncertainty. We also float $N_{osc}$, the actual (but unmeasured) number
of neutrinos in each experiment that would have been detected had they
not oscillated to a different neutrino flavor.

\FIGURE{
\centerline{\includegraphics[width=6.5 truein]{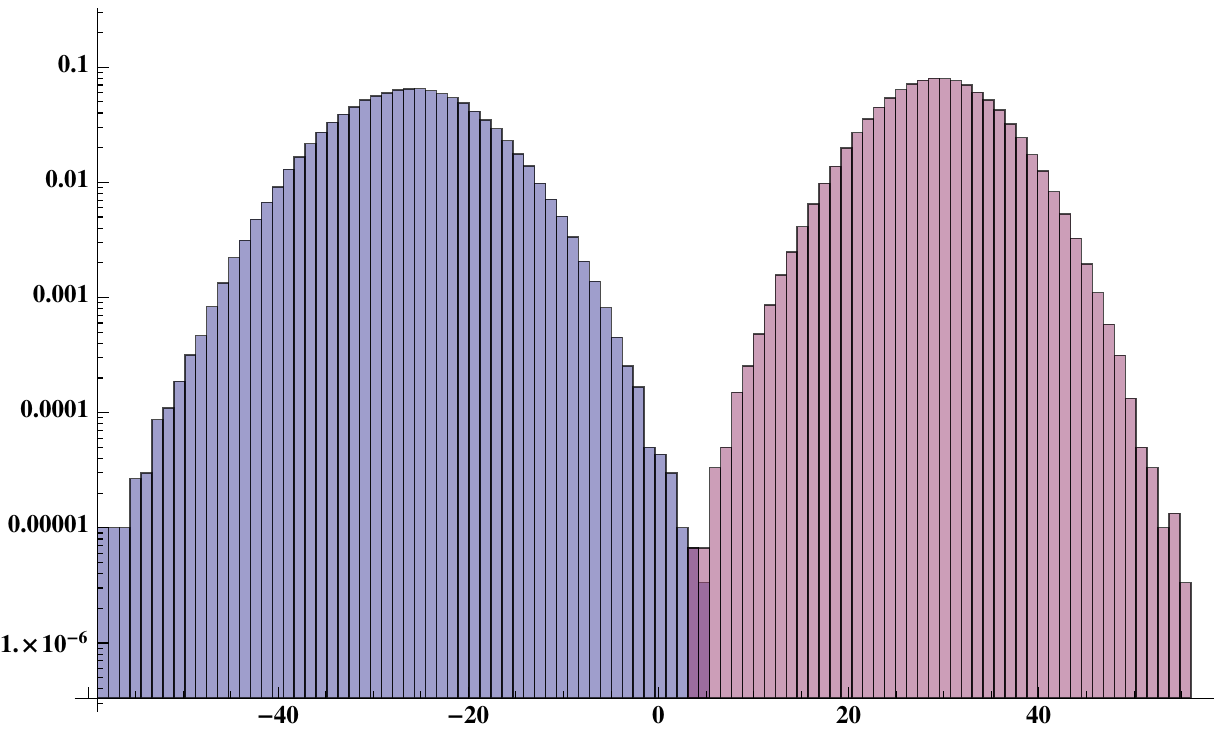}}
\caption
{\label{fig:five}
The log likelihood ratio distribution for 300,000 toy experiments simulating
200 nominal muon antineutrinos per experiment on the MINOS baseline.
The right-hand histogram uses the $CPT$-conserving hypothesis
to generate the toy results, while the left-hand histogram uses
the $CPT$-violating hypothesis. In both cases the ratio is from the
maximum likelihood computed with the $CPT$-conserving pdfs over the
maximum likelihood computed with the $CPT$-violating pdfs. The
histograms are normalized to unit probability.
}
}

Each toy experiment is equivalent to MINOS running with 200 nominal muon antineutrino
events expected in the far detector in the absence of oscillations.
This is approximately a factor of 3 increase over the current data.
For each of 300,000 toy experiments based on each mass spectrum, 
we compute the maximum likelihood (\textit{i.e.} we maximize the
likelihood with respect to the floated parameters) for both $CPT$-conserving
and $CPT$-violating hypotheses; then we plot the normalized distribution of events
versus the logarithm of the ratio of the likelihoods.  The definition of the likelihood
and the details of the analysis are presented in the appendix.
The result is shown in Figure \ref{fig:five}.

This plot allows a Neyman-Pearson test of the $CPT$-conserving versus
$CPT$-violating hypotheses. We choose a cut $\alpha$ on the log ratio
of the likelihoods for the case that the toys are based on the $CPT$-conserving 
spectrum; the significance $\alpha$ corresponds to the probability that 
the $CPT$-conserving hypothesis is rejected even though it is true.
Clearly we should choose a small value for $\alpha$, since we have a strong
prior bias that $CPT$ is conserved. Having thus fixed $\alpha$ we can extract
$\beta$, the probability that the $CPT$-conserving hypothesis is accepted
even though the $CPT$-violating spectrum is the correct one. Then
$1-\beta$ is the measure of the power of this hypothesis test. 

The results are very encouraging: even with $\alpha$ chosen as small as 
$6\times 10^{-7}$,  corresponding to a Gaussian significance of $5\sigma$,
we find $1-\beta$ very close to unity. This indicates a nearly 100\% chance 
that the $CPT$-violating spectrum discussed in this paper would be 
correctly chosen by the hypothesis test if it is in fact true.

Figure \ref{fig:six} shows the oscillation probabilities of both the
$CPT$-conserving and $CPT$-violating hypotheses plotted as a
function of the muon antineutrino energy. The current MINOS data
points (binned in energy) are superimposed.
From this figure it is clear that the $CPT$-violating hypothesis makes
clear energy-dependent predictions about what should be observed
in future MINOS running:
\begin{itemize}
\item The lowest energy bin will rise.
\item The dip apparent to the eye around 10 GeV will remain.
\end{itemize}

This figure also explains why our prediction for the power
of Neyman-Pearson test with 200 nominal MINOS events is so
encouraging, even though we have very conservatively floated the 
total number of neutrinos in the likelihood fits. The discrimination
of the $CPT$-conserving and $CPT$-violating hypotheses comes
both from a significant difference in the overall survival
probabilities and from the dramatic difference in the energy dependence of
the survival probabilities.
Even allowing for a rather large systematics, as we have done here, data generated from one
hypothesis is almost never as well-described by the
incorrect hypothesis, for experiments with at least 200 events.

\FIGURE{
\includegraphics[width=6.0 truein]{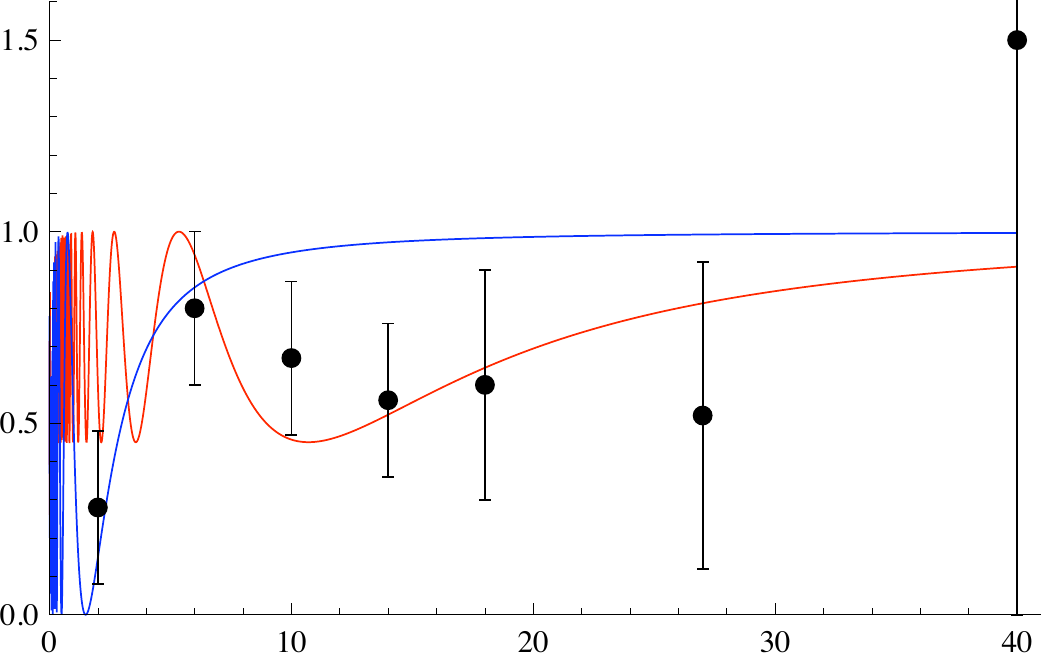}
\caption
{\label{fig:six}
Comparison of the binned MINOS muon antineutrino disappearance data
with the oscillation curves obtained from the $CPT$-conserving hypothesis
(blue) and the $CPT$-violating hypothesis (red).
}
}

\subsection*{Acknowledgments}
The authors are grateful to Milind Diwan, Georgia Karagiorgi, Bill Louis, Olga Mena, Maurizio Pierini,
Chris Quigg, and Chris Rogan.
for useful discussions. 
GB acknowledges support from the Spanish MEC and FEDER under 
Contract FPA 2008/02878, and a Prometo grant.
Fermilab is operated by the Fermi Research Alliance LLC under
contact DE-AC02-07CH11359 with the U.S. Dept. of Energy.

\appendix
\section{Likelihood methods and neutrino oscillations}

Consider a typical neutrino oscillation disappearance experiment,
in which $n_i$ neutrinos of a particular flavor are observed in a far
detector in some number of energy bins labelled by $i$. Let $N_s = \sum_i n_i$
be the total number of neutrinos observed, and for simplicity ignore the
possibility of fakes or neutrinos from background sources.

Using observations in a near detector or some other method, one
computes the mean number of neutrinos $N_0$ that one would have
expected to observe in the far detector in the absence of neutrino oscillations.
For simplicity assume that $N_0$ is the mean of a Poisson distribution,
although one could also handle more complicated distributions.
The experiment will also calculate the energy distribution of neutrinos
expected at the far detector in the absence of oscillations. Both of the
aforementioned distributions are hypotheses, perhaps with floating
parameters representing uncertainties, but in both cases assume that
the hypothetical distributions are independent of the particular neutrino oscillation
model being tested.

Now suppose one has a hypothesis for the correct neutrino oscillation model.
This amounts to specifying the pdf $p_s(E)$, the probability that a neutrino
of energy $E$ does not oscillate to a different flavor. Convolving these
pdfs with the energy distribution one obtains $p_s^i$, the probability that
a neutrino is in the $ith$ energy bin and did not oscillate. Letting
$p_s^{total} = \sum_i p_s^i$, the probability that a given neutrino does
oscillate to a different flavor is just $p_{osc}=1-p_s^{total}$.

The appropriate binned extended likelihood function given all these
assumptions is given by
\be\label{eqn:ourL}
L = \frac{{\rm e}^{-N_0} (N_0)^{N_s+N_{osc}}}{(N_s+N_{osc})!}
\;\frac{(N_s+N_{osc})!}{N_s! \;N_{osc}!}\;
p_{osc}^{N_{osc}}\;\prod_{i=1}( p_s^i )^{n_i} 
\;,
\ee
where $N_{osc}$ denotes the total number of neutrinos that oscillated
to a different flavor. 

In most applications of the extended likelihood formula, the analog
of the total number of events, here $N_s + N_{osc}$, is known, while
the mean expected number $N_0$ is obtained from the fit by maximizing the
likelihood. In the case at hand an estimate of $N_0$ is already given,
$N_s$ is measured, and $N_{osc}$ is unknown. Thus one would like to
obtain $N_{osc}$ by maximizing the likelihood. From the explicit dependence
shown in \ref{eqn:ourL}, it is easy to see that the likelihood is maximized
by solving
\be\label{eqn:maxLosc}
\psi_0(N_{osc}+1) = {\rm log}[N_0\,p_{osc}]  
\; ,
\ee
where $\psi_0$ is the digamma function.

For  $N_{osc}\gsim 2$,  an excellent approximation to the solution
of \ref{eqn:maxLosc} is given by 
\be
N_{osc} = N_0\, p_{osc}-\frac{1}{2} 
\; .
\ee
Substituting back into \ref{eqn:ourL}, we obtain an expression for
the likelihood function already maximized with respect to the floating
value of $N_{osc}$:
\be\label{eqn:ourfinalL}
L = \frac{{\rm e}^{-N_0} (N_0)^{N_s+N_0\,p_{osc}-1/2}}{N_s! \;\Gamma[N_0\,p_{osc}+\frac{1}{2}]}\;
p_{osc}^{N_0\,p_{osc}-1/2}\;\prod_{i=1}( p_s^i )^{n_i} 
\;.
\ee
This likelihood can then be further maximized with respect to other floating
parameters. 

In a real experiment the mean expected number $N_0$ is estimated from
other data with some error. If we take this error to be Gaussian,
we can include this distribution in the definition of the extended likelihood,
and maximize the likelihood with respect to both $N_0$ and $N_{osc}$.
The likelihood function becomes
\be\label{eqn:ourLb}
L = \frac{{\rm exp}\left(-\frac{(N_0-\bar{N}_0)^2}{2\sigma_{N_0}}\right) }{\sigma_{N_0}\sqrt{2\pi}}               
\frac{{\rm e}^{-N_0} (N_0)^{N_s+N_0\,p_{osc}-1/2}}{N_s! \;\Gamma[N_0\,p_{osc}+\frac{1}{2}]}\;
p_{osc}^{N_0\,p_{osc}-1/2}\;\prod_{i=1}( p_s^i )^{n_i} 
\;,
\ee
where the parameters $\bar{N}_0$ and $\sigma_{N_0}$ are supposed to be fixed
by, \textit{e.g.}, extrapolating from near detector data. In our fits we have used
$\bar{N}_0 = 200$ and $\sigma_{N_0} = 20$. We then maximize the likelihoods
allowing $N_0$ to take any value such that $N_{osc}$ is nonnegative.
This increases the
maximized likelihood of the wrong hypothesis while having very little
effect on the maximized likelihood for the correct hypothesis; thus
floating the value of $N_0$ decreases the log 
likelihood ratio of the correct hypothesis over the wrong hypothesis.

\FIGURE{
\includegraphics[width=5.0 truein]{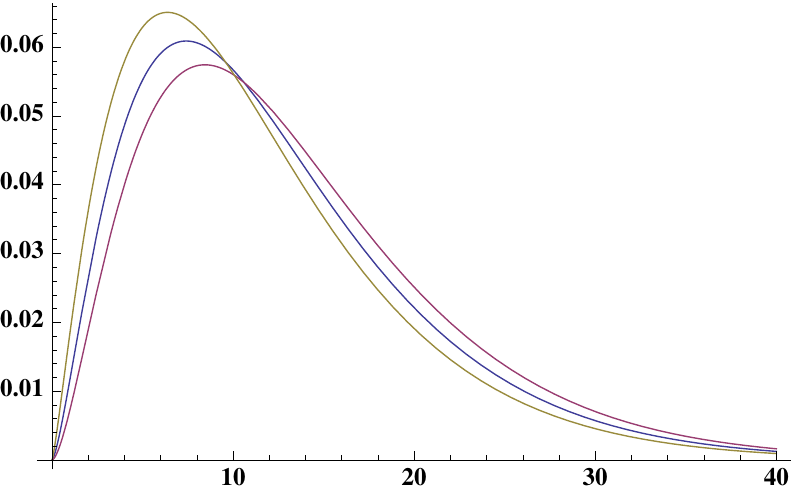}
\caption
{\label{fig:seven}
The one-parameter variability in the normalized muon antineutrino energy
spectrum floated in our maximum likelihood fits. The central (blue) curve
is our best fit to the binned MINOS spectrum. The horizontal axis
is energy in GeV.
}
}

In our analysis we introduced a parameter $c$ to represent 
uncertainty in the normalized energy distribution of neutrinos reaching the MINOS
far detector:
\be
p(E) = \frac{{\rm e}^{-aE}\,E^b}{a^{-(b+1)}\,\Gamma[b+1]}
\;,\quad a=0.193489\,;\; b= 1.43356 + c
\; ,
\ee
where the numerical constants come from a fit to the MINOS spectrum.
By allowing $c$ to vary from $-0.2$ to 0.2 in the fit independently for
each neutrino oscillation hypothesis, we introduce a variability in the
energy spectrum as illustrated in Figure \ref{fig:seven}. This increases the
maximized likelihood of the wrong hypothesis while having very little
effect on the maximized likelihood for the correct hypothesis; thus
allowing a distorted energy spectrum decreases the log 
likelihood ratio of the correct hypothesis over the wrong hypothesis.



\end{document}